\documentstyle[aps,preprint,tighten,epsf]{revtex}

\newcommand{\ket}[1]{\left|{#1}\right\rangle}
\newcommand{\nnb}{$n\bar{n}$}
\newcommand{\kbk}{$K\bar{K}$}
\newcommand{\ssb}{$s\bar{s}$}
\newcommand{\qqb}{$q\bar{q}$}
\newcommand{\pbp}{$\bar{p}p$}

\begin{document}
\title{Scalar Mesons in a Relativistic Quark Model with Instanton-Induced
Forces}
\author{E.\,Klempt\footnote{E-mail: klempt@viskpa.iskp.uni-bonn.de}$^{,1}$,
B.\,C.\,Metsch\footnote{E-mail: metsch@itkp.uni-bonn.de}$^{,2}$,
  C.\,R.\,M\"unz\footnote{E-mail: muenz@itkp.uni-bonn.de}$^{,2}$,
  H.\,R.\,Petry$^{2}$}
\address{$^{1}$Institut f\"ur Strahlen- und Kernphysik, \\
         $^{2}$Institut f\"ur Theoretische Kernphysik,\\
         Universit\"at Bonn, Nu{\ss}allee 14-16, 53115 Bonn, Germany\\
         $\,$\\
         }
\date{\today}

\preprint{\vbox{Bonn TK-95-19 \hfill Submitted to Physics Letters B}}
\maketitle

\begin{abstract}
  In a relativistic quark model with linear confinement and an
  instanton-induced interaction which solves the $\eta$-$\eta'$ puzzle, scalar
  mesons are found as almost pure SU(3) flavor states.  This suggests a new
  interpretation of the scalar nonet: We propose that the recently discovered
  $f_0(1500)$ is not a glueball but the scalar (mainly)--octet meson for which
  the $K\bar{K}$ decay mode is suppressed. The mainly--singlet state is
  tentatively identified with the $f_0(980)$. The isovector and isodoublet
  states correspond to the $a_0(1450)$ and $K^{\ast}(1430)$, respectively.
\end{abstract} \pacs{BONN TK-95-19}
\narrowtext
\section{Introduction} \label{I}

The spectrum of scalar mesons is puzzling. The number of resonances found in
the region from 1 to 2 GeV~\cite{PDG94} exceeds the number of states which can
be accommodated in conventional quark models. Extra states are interpreted
alternatively as $K\bar{K}$ molecules, glueballs, multi-quark states or
hybrids. In particular the $f_0(1500)$ \cite{cb} is presently considered as
prime glueball candidate \cite{amslerclose}. This state is seen in decays into
$\pi\pi ,\; \eta\eta ,\; \eta\eta^{\prime}$, and into '$\sigma\sigma$' (where
'$\sigma$' stands for the scalar $\pi\pi$ interaction). It has been observed in
\pbp\ annihilation at rest \cite{cb} and in flight \cite{E760}, in
Pomeron--Pomeron interactions \cite{gams,wa89} and in radiative J/$\psi$ decays
\cite{toby}.
\par
All these processes are supposed to be 'glue--rich' and enhance the chance of
observing glueballs. The mass of the $f_0(1500)$ compares very well with
predictions of lattice gauge calculations \cite{teper}. The main argument in
favor of the glueball interpretation is, however, the peculiar decay pattern.
It decays strongly into $\pi\pi$ but not into $K\bar{K}$. Assuming the
$f_0(1300)$ to be one of the two isoscalar states, the second isoscalar state
should decay preferentially into $K\bar{K}$. Hence the $f_0(1300)$ and the
$f_0(1500)$ cannot be the two isoscalar states of one meson nonet.
\par
These arguments lead naturally to the hypothesis that the $f_0(1500)$ is a
glueball. This interpretation then requires the existence of a further scalar
state which is mainly $s\bar{s}$ and should have a mass of about 1700 MeV,
possibly the old $\Theta (1690)$. The scalar meson nonet would be nearly
ideally mixed.  The peculiar decay properties of the $f_0(1500)$ can be
reproduced by tuning its mixing with the $f_0(1300)$ \nnb\ and the
(predicted) $f_0(1700)\ s\bar{s}$ state \cite{amslerclose}.  The $f_0(1500)$ is
hence interpreted as glueball state with strong mixing with close--by
conventional scalar mesons.
\par
In this paper we propose a radically different interpretation of the spectrum
of light scalar mesons. The states are interpreted as conventional $q\bar{q}$
states, but with very small SU(3) mixing angle, governed dynamically by
't Hooft's instanton-induced interaction. The comparison of the predicted mass
spectrum with data gives a surprisingly good agreement; no further states are
needed.  The reduced $K\bar{K}$ partial decay width of the $f_0(1500)$ is
qualitatively explained by its flavor structure.

\section{Experimental Scalar Meson Mass Spectrum} \label{II}
We start our discussion with a short review of the experimental situation. The
Particle Data Group \cite{PDG94} lists two scalar isodoublet states, the
$K^{\ast}_0(1430)$ and the $K^{\ast}_0(1950)$, and one isovector $a_0(980)$. A
second isovector meson has been discovered recently~\cite{amsler94}, the
a$_0(1450)$, with mass and width of $(M,\Gamma)=(1450 \pm 40\,\mbox{MeV},270
\pm 40\,\mbox{MeV})$, respectively.  There are 5 isoscalar states in the Review
of Particle Properties, the $f_0(980)$, $f_0(1300)$, $f_0(1370)$, $f_0(1525)$,
$f_0(1590)$ and 2 further possibly scalar states, the $f_J(1710)$ seen in
radiative $J/\psi$ decays (the old $\Theta (1690)$) and an $\eta$-$\eta$
resonance X(1740) produced in $p\bar{p}$ annihilation in flight and in
charge--exchange.  A rather narrow isoscalar state with $(M,\Gamma)=(1450 \pm
5\,\mbox{MeV},54 \pm 7\,\mbox{MeV})$ has been reported in \cite{wa89}; it is
produced in central collisions and seen in decays into 4$\pi$.  A recent
reanalysis of data on $J/\psi\rightarrow\gamma 2\pi^+2\pi^-$ revealed the
existence of three isoscalar states at 1505 MeV, 1750 MeV and at 2104 MeV
\cite{toby}.  These are certainly more resonances than a quark model can
accommodate.  Hence we first try to identify those which we will interpret as
\qqb\ mesons.
\par
The $K_0^{\ast}(1430)$ is the only strange candidate with scalar quantum
numbers; it gives a natural mass scale for the 1$^3P_0$ light--quark nonet.  We
use the $K^{\ast}_0(1950)$ to estimate the excitation energy of scalar radial
excitations to 520\,$\mbox{MeV}$. This may be compared to the mass difference
of 370\,$\mbox{MeV}$ between the $\chi_{b0}(2P)$ and $\chi_{b0}(1P)$ states.
\par
It has been convincingly argued, that the narrow a$_0(980)$, which has also
been seen as a narrow structure in $\eta\pi$-scattering, can be generated by
meson-meson dynamics alone~\cite{Wei90,Jan95}.  This interpretation of the
a$_0(980)$ leaves the a$_0(1450)$ as the $1^{3}P_0$ quark-antiquark state.  In
analogy, it is mostly assumed that the $f_0(980)$ is a $K\bar{K}$ molecule. The
mass degeneracy and their proximity to the $K\bar{K}$ threshold seem to require
that the nature of both states must be the same. On the other hand, the
$K\bar{K}$ interaction in isospin I=1 and I=0 are very different \cite{speth}.
The extremely attractive I=0 interaction may not support a loosely bound state.
Instead, it may just define the pole position of the $f_0(980)$ $q\bar{q}$
resonance. Indeed, Morgan and Pennington find a $f_0(980)$ pole structure
characteristic for a genuine resonance of the constituents and not of a weakly
bound system \cite{Mor93a,Mor93b}.  The I=1 $K\bar{K}$ interaction is weak and
may generate a $K\bar{K}$ molecule.
\par
The next two states, the f$_0(1300)$ and f$_0(1370)$, deserve particular
attention. We follow the arguments of~\cite{Mor93a,Mor93b} and assume that the
$\pi\pi$ interactions produce a very broad f$_0(1000)$ state, and a
comparatively narrow f$_0(980)$ giving rise to the dip at 980 MeV in the
squared $\pi\pi$ scattering amplitude $T_{11}$. In this scenario the
f$_0(1300)$ is interpreted as the high-mass part of the f$_0(1000)$. In
experiments the $f_0(1000)$ shows up as a resonance at 1300 MeV because of the
pronounced dip in $|T_{11}|^2$ at 1 GeV.  The $f_0(1000)$ has an extremely
large width; thus the resonance interpretation is questionable. It could be
generated by t--channel exchanges instead of inter--quark forces.  As for the
second state, we do not consider the scalar 4$\pi$ resonance seen in
$N\overline{N}$ annihilation into 5$\pi$ by Gaspero \cite{gaspero}, by the
Obelix \cite{adamo} and the Crystal Barrel \cite{curtis} collaborations as
established. Likely, the mass of it is compatible with 1500\,$\mbox{MeV}$
\cite{resag}.
\par
Finally, we notice several claims for resonant structures close to
$1500\,\mbox{MeV}$ including the states f$_0(1450)$, f$_0(1500)$ and
$f_0(1590)$. Their masses, widths and decay branching ratios are incompatible
within the errors quoted by the groups. Nevertheless, we do not consider it as
plausible, that so many scalar isoscalar states exist in such a narrow mass
gap. Instead we take the various states as manifestations of one object which
we call f$_0(1500)$.
\par
These arguments lead us to the suggestion to identify the following states as
members of the ground state scalar \qqb\ nonet:
\begin{equation}
a_0(1450),\;\;\;\;\;\; K_0^*(1430),\;\;\;\;\;\; f_0(980),\;\;\;\;\;\;
f_0(1500)
\label{ident}
\end{equation}
In view of the huge mass splitting a grouping of these states into the flavor
nonet at first sight does not seem very obvious. In addition a quark model
interpretation for the isoscalar resonance at 1500 MeV meets serious
difficulties, because it is expected to be an $s\bar{s}$ state and hence
strongly decay into $K\bar{K}$. This is not observed and consequently it is
considered as a serious candidate for a glueball. We will argue that both
issues are in fact closely related.
\par
If the $f_0(1500)$ is interpreted as a $q\bar{q}$ state, it must have a special
structure, which suppresses the decay into kaons. In~\cite{amslerclose}, it is
interpreted as glueball mixing with the f$_0(1300)$ and one further state at a
mass of 1600 - 1800 MeV which is mostly \ssb. In our scenario the $f_0(1500)$
has an $s\bar{s}$ and an $n\bar{n}$ component, with a negative sign
suppressing $K\bar{K}$ decays by destructive interference. We are in fact not
unfamiliar with such a situation: it occurs already for the pseudoscalars in
the form of the $\eta$ and the $\eta'$ splitting and mixing.  Also the order of
magnitude of the mass splitting is similar.  However, for pseudoscalars the
state with a negative sign between the strange and nonstrange components (i.e.
the $\eta$-meson) is the lowest state. In the next section, we will present a
quark model which relates scalar and pseudoscalar mesons in more detail and
could offer in fact a consistent explanation of the scalar mesons without
introducing additional yet unobserved states as in~\cite{amslerclose}.

\section{A relativistic quark model of mesons with Instanton-Induced Forces}
\label{III}

The model is formally based on the Bethe-Salpeter equation for quark-antiquark
amplitudes, with free quark propagators with an effective constituent quark
mass, an instantaneous interaction kernel that models confinement and a
residual interaction based on instanton effects~\cite{Hoo76,SVZ80}. In such a
framework the mass spectrum is obtained from the equation
\begin{eqnarray}
  \Phi_P(\vec p) &=& \int\!  dp^0\; S^F_1(P/2+p)\, \int\! \frac{d^3
    p'}{(2\pi)^4}\, [-i\,V(\vec p,\vec{p}\, ') \,\Phi_P(\vec{p}\, ')]\,
  S^F_2(-P/2+p)
\end{eqnarray}
where in the CM-frame $P=(M,\vec 0)$ and \(S^F_i(p) = i/(p\!\!
/-m_i+i\epsilon)\) with a constituent-quark mass $m_n=306\,\mbox{MeV}$
and $m_s=503 \,\mbox{MeV}$ for nonstrange and strange quarks,
respectively. The interaction kernel $V$ comprises a confinement part
$V_C$:
\begin{equation}
  \left[V_C(\vec{p},\vec{p}\,')\,\Phi(\vec{p}\,')\right] = {\cal
    V}_C((\vec{p}-\vec{p}\,')^2)\;
  \frac{1}{2}
  \left[\Phi(\vec{p}\,')-\gamma^0\,\Phi(\vec{p}\,')\,\gamma^0\right]
\label{conf}
\end{equation}
where the scalar function \({\cal V}_C\) in coordinate space is given by a
linearly rising potential \({\cal V}_C(r) = a_c+b_c r\), with
$a_c=-1751\,\mbox{MeV}$ and $b_c=2076\,\mbox{MeV}$.  The special Dirac
structure of $V_C$ was chosen in order to obtain a maximal cancelation of
unwanted spin-orbit splitting.  The quark masses and confinement parameters
have been fixed to reproduce the overall meson spectrum except for the mesons
with spin zero. We find a very good description of the meson masses, especially
for the Regge-behavior, which will be presented in a detailed
study~\cite{HMMP}.

In addition the kernel contains a residual interaction $V_T$ in the form of 't
Hooft's instanton-induced interaction:
\begin{eqnarray}
  [V_T(\vec{p},\vec{p}\,')\,\Phi(\vec{p}\,')] = 4\,\,G \left[
  1\,\mbox{tr}\,\left(\Phi(\vec{p}\,')\right) +
  \gamma^5\,\mbox{tr}\,\left(\Phi(\vec{p}\,')\,\gamma^5\right)\,\right]
\label{thokern} \end{eqnarray}
where $G(g,g')$ is a flavor matrix containing the coupling constants
\mbox{$g=13.2\,\mbox{MeV\,fm$^3$}$} and $g'=11.8\,\mbox{MeV\,fm$^3$}$, which
multiply the $n\bar{n}$- and $n\bar{s}$-interaction, respectively. Here
summation over flavor and a regulating Gaussian function with a range
$\lambda=0.3 \mbox{ fm}$ have been suppressed (see \cite{Mue94} for more
details).  The parameters of the 't Hooft interaction have been fixed in order
to reproduce the spectrum of the pseudoscalar states, as this force has been
suggested to describe the breaking of the U$_A$(1) symmetry.

As it stands, with this instantaneous interaction kernel the Bethe-Salpeter
equation is reduced to the three dimensional Salpeter equation, which can be
solved by standard techniques: all details can be found
elsewhere~\cite{Mue94,RMMP94}, together with a discussion of the relativistic
quark amplitudes and their behavior under Lorentz-boosts, which improves
dramatically the values computed for form factors and electroweak decay values
of deeply bound quark states in comparison to nonrelativistic quark
models~\cite{Mue94b}.

\begin{figure}[htb]
  \begin{center}
  \leavevmode
  \epsfxsize=0.65\textwidth
  \epsffile{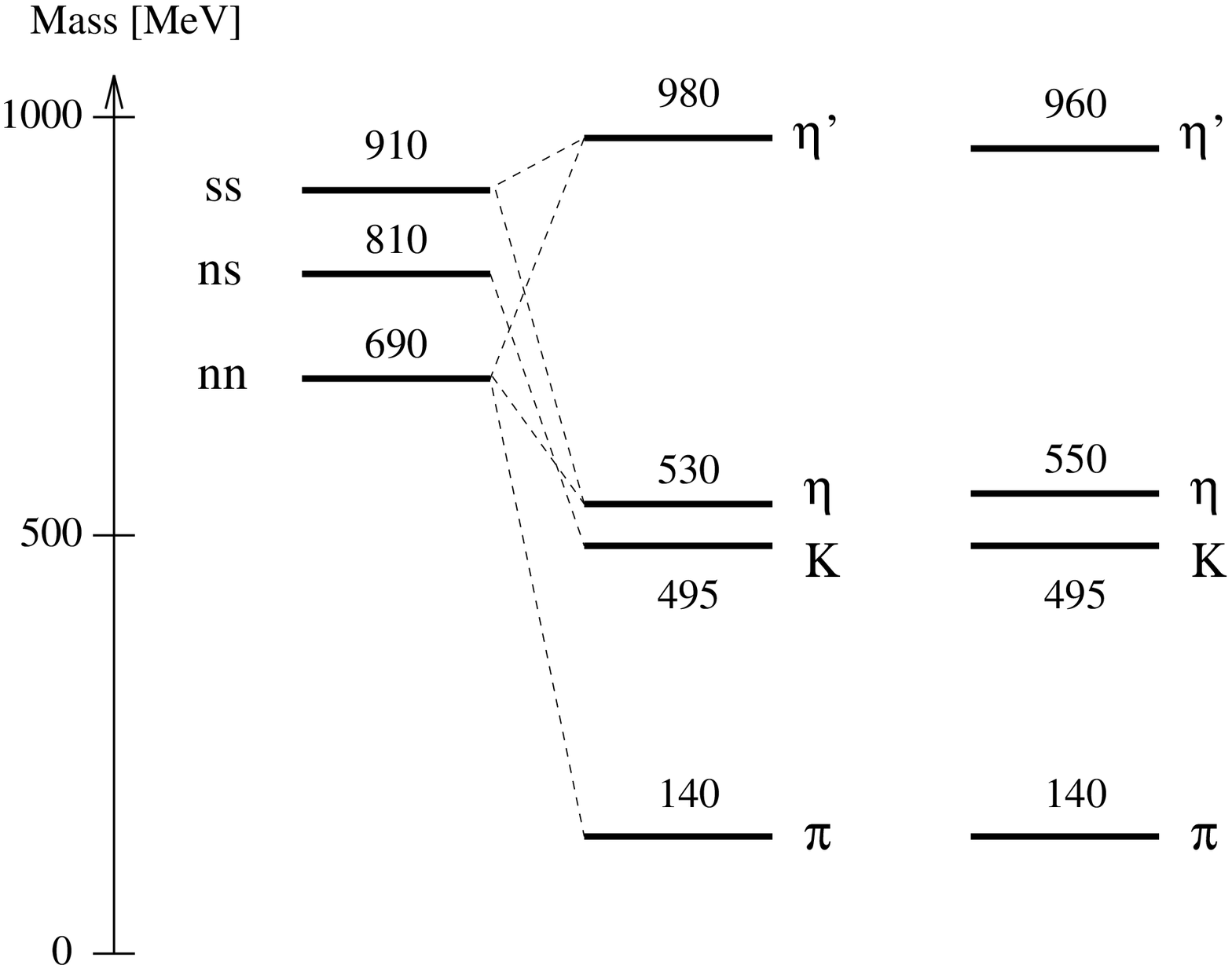}\vspace{0.5cm}
  \caption{
    Schematic splitting of pseudoscalar flavor nonets with confinement
    interaction (left), with confinement and instanton-induced force (middle)
    compared to the compilation by the Particle Data Group
    \protect{\cite{PDG94}} (right).  }
  \label{Abb1}
  \end{center}
\end{figure}

In fact we have used the same instanton-induced interaction before in a
nonrelativistic reduction within the framework of the Schr\"odinger equation,
both for mesons and baryons~\cite{Bla90} and obtained a satisfactory
description of the low lying hadronic mass spectrum. In particular 't Hooft's
interaction naturally explained (consistently with the major splitting in the
baryon spectrum) the splitting and configuration mixing of the pseudoscalar
nonet. This feature remains in the present relativistic treatment, see
Fig.~\ref{Abb1}.  However, in the nonrelativistic reduction, which is
essentially an expansion in average quark momenta divided by the constituent
quark mass (a quantity of the order one!), 't Hooft's force vanishes for scalar
mesons.

\begin{figure}[htb]
  \begin{center}
  \leavevmode
  \epsfxsize=0.65\textwidth
  \epsffile{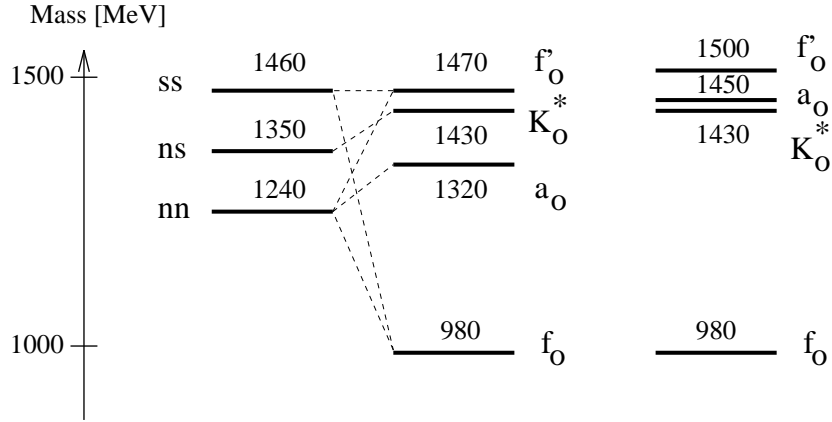}\vspace{0.5cm}
  \caption{
    Schematic splitting of pseudoscalar flavor nonets with
    confinement interaction (left), with confinement and instanton-induced
    force (middle) compared to the experimental
    spectrum interpreted as \qqb\ states~\protect{\cite{PDG94,cb}}
    (right).
          }
  \label{Abb2}
  \end{center}
\end{figure}

In the present relativistic treatment the 't Hooft interaction for scalars is
in fact appreciable: More quantitatively, it is of the same magnitude but
opposite in sign in comparison to the pseudoscalar mesons. Thus it lowers
states whose SU(3) flavor structure is dominantly flavor singlet and pushes the
dominantly flavor octet states to higher masses, see Fig.~\ref{Abb2}.  The
calculated masses of the ground state scalar mesons are predicted to be
\begin{equation}
a_0(1370),\;\;\;\;\;\; K_0^*(1430),\;\;\;\;\;\; f_0(980),\;\;\;\;\;\; f_0(1470)
\end{equation}
which is in fair agreement with (\ref{ident}), especially in view of the fact,
that the parameters were chosen to reproduce the spectrum in other meson
sectors.

It is interesting to compare the full low--lying spectra of the scalar and
pseudoscalar nonets in some more detail.  Figs.~\ref{Abb1},\ref{Abb2} show that
the mass splitting in the scalar octet is reduced. This is reasonable since the
average relative momenta should be larger for orbital excitations, and thus the
different quark masses breaking the SU(3) symmetry play a less important role.
Also the flavor splitting between singlet and octet is much more pronounced
than in the corresponding pseudoscalar case.

On the basis of the pure quark model it is hard to decide, whether to identify
the low lying mainly singlet state with $f_0(980)$ or with $f_0(1000)$.
Ultimately, this can be decided only from a genuine coupled channel approach,
which includes both \qqb\ and meson-meson states. For the time being, we {\em
  tentatively} prefer the identification with $f_0(980)$, since at least one of
the states of the nonet should couple to \kbk .

The decay properties of the scalar mesons certainly depend most sensitively on
the ratio of the \nnb\ and \ssb\ component of the wave function. Following the
parameterization of Rosner~\cite{Rosn}, we define
\begin{eqnarray}
   \ket{f_0} & = & X_{f_0} \ket{n\bar{n}} + Y_{f_0} \ket{s\bar{s}}
\\
   \ket{f'_0} & = & X_{f'_0} \ket{n\bar{n}} + Y_{f'_0} \ket{s\bar{s}} \nonumber
\end{eqnarray}
Although the radial wave functions of the nonstrange and strange component in
our model are different, we have calculated the value of the coefficients $X$
and $Y$ from the relativistic norm~\cite{Mue94} and the apparent relative sign
of the amplitudes. The values are given in Table~\ref{tabmix} and indicate,
that due to the instanton-induced interaction the $f_0$ is almost a SU(3)
flavor singlet and the $f'_0$ is almost an octet.

The scalar mesons decay electro-magnetically into two photons as well as into a
photon and a vector meson.  Emission of photons is certainly an important
test--ground to decide on the glueball-- or $q\bar{q}$--nature of the
$f_0(1500)$ since gluons do not couple to photons. Corresponding $\gamma\gamma$
results in the tensor sector agree very favorably with data, which indicates
the reliability of the calculations. The results will be published in a
forthcoming paper~\cite{Mue95}.

The two--photon and the photon--vector-meson widths have been calculated in the
framework of the Salpeter model following the relativistic procedure outlined
in~\cite{RMMP94,Mue94b} including e.g.\ explicitly the negative energy
components. The results on scalar mesons are shown in Table~\ref{tabgg}, where
we compare results with and without the instanton-induced interaction. The
scarce experimental data are also listed.  Again we will refrain from an
ultimate experimental identification of the $f_0$ state, although the large
calculated $\gamma\gamma$-width better corresponds to the result found for
the $f_0(1300)$, which we quoted as $f_0(1000)$.  The decrease of the
$\gamma\gamma$ width of the lowest $f_0$ is merely due to the decrease of phase
space, whereas the decrease of the $\gamma\gamma$ width of the $f'_0(1500)$
comes from the destructive interference between the \nnb\ and \ssb\ component.
Unfortunately, the data are too poor and do not allow to distinguish the two
models.

The radiative decays into a vector meson again are mainly determined by phase
space except for the $f'_0(1500)$, where the results are very sensitive to the
flavor structure of this meson. In particular the decays into $\rho\gamma$ and
$\omega\gamma$ are sensitive to the \nnb , the $\Phi$ to the \ssb\ component of
the wave function.

An outstanding feature of the scalar mesons is their peculiar strong decay
pattern.  As a full calculation of hadronic decay amplitudes in the Salpeter
formalism is tuff and still under investigation, we will merely quote the
calculated mixing angle of the scalar states, which can be used to estimate the
branching ratios, see~\cite{amslerclose}. The mixing of the singlet and octet
$f_0$ is parameterized as
\begin{eqnarray}
   \ket{f_0} & = & \sin(\Theta_S) \ket{f_{0,8}}
                  +\cos(\Theta_S) \ket{f_{0,1}}
\\
   \ket{f'_0} & = & \cos(\Theta_S) \ket{f_{0,8}}
                  -\sin(\Theta_S) \ket{f_{0,1}} \nonumber
\end{eqnarray}
and use the relative \nnb\ and \ssb\ amplitudes given in Table~\ref{tabmix} to
estimate $\Theta_S$. The admixture of the \nnb\ component leads to a mixing
angle of approximately $\Theta_S=6^o$ for $f_0'$, which already decreases the
\kbk\ amplitude by a factor of 4 compared to a purely \ssb\ state
($\Theta_S=35.3^o$), while increasing the
$\pi\pi$-amplitude~\cite{amslerclose}.

We therefore stress that the instanton interaction naturally leads to a mixing
of the \nnb\ and \ssb\ component of the $f'_0(1500)$, with the tendency to
suppress \kbk\ decays, although the effect quantitatively does not suffice to
describe the unusual decay pattern of this meson.  Therefore more theoretical
effort is needed in order to arrive at a quantitative quark model prediction.
We also encourage experiments, which quantify the partial width especially into
\kbk .  In addition, a measurement the decay properties of the $a_0(1450)$
could decide, whether it is the isovector partner of the $f_0(1300)$ or the
$f_0(1500)$, and whether our interpretation of the scalar meson nonet is
correct.

\section{Conclusion} \label{V}
We have presented a relativistic model of mesons, which reproduces the
meson mass spectrum and sheds new light on the structure of the scalar meson
nonet. The $a_0(980)$ is interpreted as $K\bar{K}$ molecule or
threshold effect; there are two isoscalar resonances, the narrow $f_0(980)$
and the broad $f_0(1000)$. One of is supposed to form the flavor--singlet
state of the scalar nonet while the $f_0(1500)$ is considered as flavor
octet state. We prefer to interpret the $f_0(980)$ as $q\bar{q}$ state,
because of its strong coupling to $\bar{K}K$. The scalar mesons are
governed dynamically by 't Hooft's instanton-induced
interaction. This force, which solved naturally the $\eta$-$\eta'$ puzzle, thus
also explains quantitatively the unusual pattern of the scalar mesons
consisting of an almost SU(3) octet at about 1400 MeV and a low lying SU(3)
singlet at 1000 MeV.

We presented results for the modification of the two photon widths coming from
the flavor mixing due to the instanton force, which may serve for an
experimental verification of our model. In addition, this concept has to be
further explored in the description of the strong decay widths, for which work
is in progress.

The identification of the calculated states with experimental data on the
basis of the present model is not yet conclusive: Nevertheless, we do believe,
that the scalar particles, very much like the pseudoscalars, exhibit large
splitting and mixing properties reflecting instanton effects, and that this
should be considered also when invoking other mechanisms such as mixing with
glueballs and multi-meson states.

{\bf Acknowledgments:} We thank C.\,Amsler, D.\,V.\,Bugg, F.\,E.\,Close,
K.\,Holinde, L.\,Montanet, D.\,Morgan, J.\,M.\,Richard and J.\,Speth for
helpful discussions. Early contributions by S.\,Hainzl are acknowledged.
M.\,R.\,Pennington took part in a workshop on scalar resonances at the
University of Bonn in April 1995. His contributions are particularly
appreciated.


\begin{table}
  \caption{Calculated \protect{$\gamma\gamma$}-widths in eV with and
    without instanton-induced interaction compared to experimental
    data~\protect{\cite{PDG94}} }
  \label{tabgg}
  \centering
   \begin{tabular}{cccc}
        & full calculation& confinement only &  experimental \cite{PDG94}\\
     \hline
       $f_0(980)\rightarrow\gamma\gamma$ & 1810 & 3400 & 560 $\pm$ 110
                                                       ($f_0(980)$)  \\
                                         &      &      & 5400 $\pm$ 2300
                                                       ($f_0(1300)$)  \\
       $f_0(1500)\rightarrow\gamma\gamma$ & 155  & 280  &  \\
       $a_0(1450)\rightarrow\gamma\gamma$ & 1370 & 1230 &  \\
$f_0(980)\rightarrow\rho\gamma$ & 14.9 & 190 &    \\
$f_0(980)\rightarrow\omega\gamma$ & 1.57 & 20 &    \\
$f'_0(1500)\rightarrow\rho\gamma$ & 160 & 0 &    \\
$f'_0(1500)\rightarrow\omega\gamma$ & 17.4 & 0 &    \\
$f'_0(1500)\rightarrow\Phi\gamma$ & 87 & 101 &    \\
$a_0(1450)\rightarrow\rho\gamma$ & 34 & 21 &    \\
$a_0(1450)\rightarrow\omega\gamma$ & 290 & 180 &    \\
$K_0(1430)\rightarrow K^*\gamma$ & 190 & 124 &    \\
\end{tabular}
\end{table}

\begin{table}
  \caption{\(f_0,f'_0\) mixing parameters from relativistic norm}
  \label{tabmix}
  \centering
   \begin{tabular}{ccc}
   Meson &  X & Y   \\
\hline
   \(f_0 (980)\) &  $\;$0.92 & $\;\;\,$0.40\\
   \(f'_0 (1500)\) &  $\;$0.48 & $-$0.88
\end{tabular}
\end{table}

\end{document}